\documentclass{nature}
\usepackage{graphicx}
\usepackage{amsmath}
\usepackage{amssymb}

\title{Analog and digital phase modulation of spin torque nano-oscillators}

\author{A. Litvinenko$^1$, P. Sethi$^1$, C. Murapaka$^1$, A. Jenkins$^2$, V. Cros$^3$, P. Bortolotti$^4$, R. Ferreira$^2$, B. Dieny$^1$ \& U. Ebels$^1$}

\begin{document}

\maketitle

\begin{affiliations}
 \item Univ. Grenoble Alpes, CEA, CNRS, IRIG-SPINTEC, F-38000 Grenoble, France
 \item International Iberian Nanotechnology Laboratory (INL), Braga, Portugal
 \item Unité Mixte de Physique CNRS, Thales, Univ. Paris-Sud, Université Paris-Saclay, Palaiseau, France
 \item THALES TRT, Palaiseau, France
\end{affiliations}

\begin{abstract}
Spin torque nano-oscillators (STNO) are nanoscale and multifunctional devices with wide band frequency tunability. Their RF properties are well suited to define novel schemes for wireless communication used within wireless sensor networks (WSN). To meet constraints of low power consumption and autonomy, WSNs use basic protocols for data transmission which are amplitude, frequency and phase shift keying (ASK, FSK, PSK). ASK and FSK have been demonstrated for STNOs but implementation of PSK has been left open so far. Phase shift keying requires a stable phase and a well-defined reference phase, while in free running STNOs the phase is free and characterized by large fluctuations. Here we introduce a special PSK technique for STNOs by combining their modulation and injection locking functionality. Injection locking provides a stable reference phase and reduces the STNO phase noise making the system compatible with IEEE standards. We used injection locking at 2f and f/2. The STNO intrinsic frequency is detuned via injecting additional modulation signal resulting in a well-defined phase shift in the injection locked state at the constant frequency multiple to the external source frequency. The concept is validated using magnetic tunnel junction based vortex STNOs. For injection locking at 2f and f/2 we demonstrate PSK data transmission rates up to 4Mbit/s as well as the possibility of quadrature phase shift keying (QPSK). Finally we have built a bench-top transmitter-receiver system to demonstrate data transmission over a 10 meter distance using analog phase modulation of STNOs and a conventional software defined radio receiver. We discuss the steps to be taken for integration of this scheme. This demonstrates that performances of the STNO-based PSK modulator (e.g. phase noise and output power) are compatible with existing technologies and that STNO multifunctional properties can be exploited to define compact approaches for wireless communication in sensor networks. 
\end{abstract}

\section{\label{sec:level1}Introduction}
Wireless sensor networks (WSN) \cite{YangWSN2013} are amongst the key enabling technologies of the Internet of Things (IoT)\cite{Uckelmann2011}. In a WSN, a large number of nodes are interconnected and are communicating with each other. Such applications require solutions which are wideband, compact, low cost and autonomous. Recent advances in the field of spintronics\cite{Dumas2014RecentAdv, Chen2016SpinTorqueAS} led to active developments of communication systems\cite{Choi2014,Oh2014,Sharma2015ComSys} based on spin-torque nano-oscillators (STNO) using basic modulation techniques \cite{Choi2014, Purbawati2016EnhancedFM}. Spin-torque nano-oscillators (STNOs) are potential candidates to be used as rf sources within transmitter-receiver blocks due to their large frequency tunability, nanoscale size and and multifunctionality. In an STNO, a spin-polarized DC current injected through the magnetic stack induces steady-state oscillations of the free layer magnetization by compensating damping torque \cite{Slonczewski1996}. STNOs are non-isochronous, i.e. their frequency depends on the amplitude of oscillation. This dependence provides tuning of the oscillation frequency via changing the input DC current. Applying an additional RF signal (current or field) at the STNO input makes it possible to either injection lock the STNO to an external signal source \cite{Lebrun2015} ($f_{source}=Nf_{STNO}$) or to modulate the amplitude and frequency of the signal ($f_{source}<f_{STNO}$). So far only frequency \cite{Pufall2005FM, Muduli2010NonlinFMandAM, Manfrini2011, Purbawati2016EnhancedFM, Sharma2017SBModulation, Ruiz-Calaforra2017, Zahedinejad2017CurrentAMFM} and amplitude modulation \cite{Muduli2010NonlinFMandAM, Choi2014, Sharma2015ComSys, Zahedinejad2017CurrentAMFM} have been demonstrated while for data transmission there is a third basic concept which is phase modulation (PM)\cite{Haykin1988, Anderson2013PSK}, including analog PM and digital phase shift keying (PSK). PSK modulation is more bandwidth efficient than FSK modulation and has a lower bit error rate (BER) than ASK modulation\cite{Tewari2016}. However, neither analog PM nor digital PSK has been demonstrated for STNOs. The nanoscale dimension of STNOs comes at an expense of larger phase noise which leads to large phase drifts on longer timescales. The phase of the STNO is free and, therefore, fluctuations of the phase can accumulate over time. For instance, even for vortex nano-oscillators of very low linewidth, the phase drift on a $\mu s$ time scale is of the order of tens of radians \cite{Lebrun2015, Ruiz-Calaforra2017}. According to the IEEE standard for the binary PSK (BPSK) \cite{IEEE2003}, maximal efficiency and noise resistance are achieved when the difference of two phase states is $\pi$. Therefore, the phase noise should be significantly reduced from tens of radians to at least $\pi$ in order to distinguish between data phase states and implement the PSK technique with STNOs. One way around this would be to use the STNO within a phase-locked loop (PLL). Such PLL operation has recently been demonstrated for STNOs of vortex or uniform magnetized states \cite{Kreissig2017, Tamaru2016}. However, the problem with PLL is that the bandwidth of modulation and with this the maximum achievable data rate is limited by the settling time and the loop filter of the PLL. Another possibility to reduce phase noise and through this to stabilize the phase dynamics on short and longer timescales, is to injection lock the STNO to an external RF current or field source \cite{Adler1946, Kiselev2003, Grollier2006, Tiberkevich2009, Demidov2014ExtSynchro, Ruiz-Calaforra2017, Tortarolo2018}. The aim of this work is to introduce and validate the principle of analog PM and digital PSK for STNOs by combining injection locking and modulation. Within applications the external RF signal source can then be replaced by a PLL-stabilized STNO.

\section{\label{sec:level2}Principle of PM and PSK for STNOs}

The idea of the PSK technique for STNOs is the following: Injection locking reduces the phase noise and provides a stable reference phase for the STNO. This is necessary to well distinguish different phase levels at reasonable modulation indices. Perfect locking has been demonstrated for MTJ based vortex STNOs when the locking ratio $N=f_{source}/f_{STNO}$ is $0.5, 1, 2$ \cite{Lebrun2015}. When injection locked, the oscillator phase ${\phi}_{STNO}$ has a finite and constant phase difference with respect to the phase of the source ${\phi}_{source}$ \cite{Zhou2007,Rippard2005,Rosenblum2001,Slavin2009,Balanov2009, Pikovsky2001} given by $\Psi={\phi}_{source}-N{\phi}_{STNO}$, with $d\Psi/dt=0$. This phase difference has two contributions. A constant one $\Psi_0$ that is determined by the coupling mechanism between the oscillator and the source (i.e. dissipative or conservative coupling as well as non-linear frequency amplitude coupling)\cite{Osipov2007}. The second contribution is determined by the detuning $\delta$, which is the difference in the free-running frequency of the oscillator and the signal source, $\delta=f_{source}-Nf_{STNO_{0}}$. It is generally accepted to consider only positive values of the detuning and define the locking range as the maximum positive detuning of the oscillator's intrinsic frequency: $\Delta\Omega=\delta_{max}/N$. Note that the value of $\Delta\Omega$ varies with the order $N$ of synchronization. In order to use the full potential of synchronization phenomena we employ both positive and negative values of the frequency detuning $\delta$ and therefore we define the full locking range as $2\Delta\Omega$. With this the phase difference as seen from the external source is given by \cite{Balanov2009,Pikovsky2001,Razavi2004}
\begin{equation}
\Psi|_{source} = 
    \begin{cases}
    \Psi_0 + arcsin(\frac{\delta}{\Delta\Omega N}), & \quad \text{at harmonics}\\
    \Psi_0 + N arcsin(\frac{\delta}{\Delta\Omega N}), & \quad \text{at sub-harmonics}
    \end{cases}
    \label{eq:1}
\end{equation}

Here, we have taken into account the possibility of injection locking at harmonics with $N=1, 2, 3,...$ and fractional injection locking at sub-harmonics\cite{Urazhdin2010} with $N=1/2, 1/4,...$. Different mechanisms\cite{Balanov2009} are responsible for harmonic and sub-harmonic synchronization which leads to the additional pre-factor $N$ for the phase difference for synchronization at sub-harmonics.

For data transmission using PSK we are interested in the phase difference that can be induced in the STNO signal since it will serve as the carrier to be phase-modulated. Moreover, for PSK we need to have a constant carrier frequency. Therefore, we keep the frequency of the external source $f_{source}$ unchanged and shift the STNO phase by detuning its intrinsic free running frequency $f_{STNO_{0}}$ via a modulation signal. Taking into account the ratio between the external and forced frequency of the STNO $f_{source}=Nf_{STNO}$ the equation (1) can be rewritten for the phase difference of the detuned and synchronized STNO:
\begin{equation} \label{eq:2}
\Psi|_{STNO}=\frac{1}{N}\Psi|_{source},
\end{equation}

From Eq. 1 and 2 and for the maximum locking range with $\delta=\Delta\Omega$, the corresponding maximum phase difference is (i) constant $\Delta \Psi_{max}=\Psi|_{STNO}(+\delta_{max})-\Psi|_{STNO}$ $(-\delta_{max})=\pi$ for $N=1, 1/2, 1/4, ...$ and (ii) $\Delta \Psi_{max}=\pi/N$ for $N=2, 3, 4, ...$.

Based on this possibility to control the STNO phase via the detuning, phase modulation of STNOs can be achieved by combining injection locking and modulation (see Fig. 1): (i) first a DC current source $I_{DC}$ excites the auto-oscillation at the carrier frequency $f_{STNO_{0}}$; (ii) an RF signal $V_{source}$ at frequency $f_{source}=Nf_{STNO}$ is used to injection lock the STNO, whereby the STNO will oscillate at the frequency $f_{STNO}$, which is multiple to external source frequency $f_{source}$ at a phase difference $\Psi|_{STNO}=\Psi_0/N$; (iii) an RF modulation signal $V_{mod}$ detunes the STNO free running frequency and hence the frequency detuning $\delta(V_{mod})$. This results in a phase modulation of $\Psi|_{STNO}=\Psi_0/N+[arcsin(\delta(V_{mod})/{\Delta\Omega N})]/N$ in case of synchronization at harmonics and $\Psi|_{STNO}=\Psi_0/N+arcsin(\delta(V_{mod})/{\Delta\Omega N})$ for sub-harmonic synchronization. Employing the principle of phase detuning, we demonstrate here analog phase modulation PM as well as digital binary phase shift keying BPSK, up to modulation frequencies of $f_{mod}=2MHz$ for MTJ based vortex STNOs. We also demonstrate the possibility of quadrature phase shift keying QPSK and voice transmission.
\clearpage
\begin{figure}
    \begin{center}
    \includegraphics[width=0.5\linewidth]{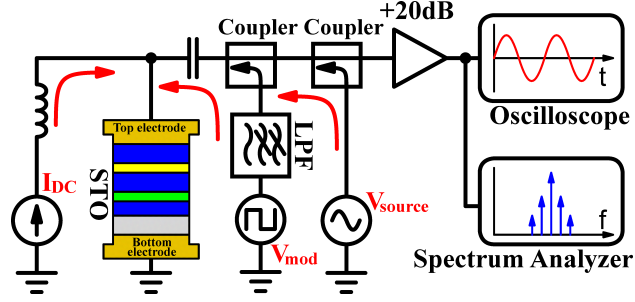}\\
    \caption{\textbf{Schematic of the electrical setup for PSK with STNOs.}}
    \label{fig1}
    \end{center}
\end{figure}

\section{\label{sec:level3}Experiments and Results}
For the experiments MTJ based vortex nano-oscillators were chosen, since they have demonstrated the capability of perfect locking to an external source \cite{Lebrun2015}. The same principle should apply for any STNO configuration, given that perfect synchronization can be achieved. The MTJ vortex STNOs used here, were realized at INL (more details see Methods). For these devices best signal generation and injection locking were obtained for nanopillar diameters of $300-400nm$ with corresponding frequencies of $[290-310]MHz$. 

A DC source (Keithley 2401) was used for signal generation, a signal generator (Agilent E8257D) for injection locking and a waveform generator (Tektronix AWG4162) for modulation. RF signals were injected via couplers to separate the STNO input and output signal. The output signal is passed through a filter (275-325MHz) and amplified by 20dB using a commercial amplifier.  The STNO output voltage signal is measured by a spectrum analyzer to deduce the carrier frequency $f_{STNO}$, linewidth ${\Delta f}$ and the power of the free running and injection-locked states as well as by a real-time oscilloscope with a single-shot mode to register time traces. From the Hilbert transform of the time traces the phase fluctuations and the phase noise properties are deduced in the free running, injection locked and phase modulated state similar to previous studies \cite{Quinsat2014, Lebrun2015, GrimaldiPhysRevB.89.104404, Ruiz-Calaforra2017, Purbawati2016EnhancedFM}. 

For the injection locking as well as PM and PSK, the operating point has been chosen such that the free running oscillator frequency is $f_{STNO_{0}}=300MHz$, which requires a DC current of $I\approx9.5mA$ and an out-of-plane field of $H\approx2.5kOe$ produced by a permanent magnet. At this operating point the free running linewidth and the power are $250kHz$ and $0.25{\mu}W$, correspondingly.The dependence of the free running oscillator frequency on the DC current can be found in the supplementary material. Then an RF sinusoidal signal at a frequency of $f_{source}= 2f_{STNO}=600MHz$ is injected at a power level of -3dBm. The frequency $f_{source}$ was kept constant and the DC current was varied. Fig. 2 (a) shows an example of the injection locking of the STNO upon varying the intrinsic frequency of the STNO via the DC current. The STNO remains locked over a locking range of DC input current ${2\Delta I}\approx0.5mA$ which corresponds to the locking range of intrinsic frequency ${2\Delta\Omega(I)}$. The corresponding linewidth \cite{Kim2008}, which is limited by the resolution bandwidth ${\Delta}f_{sa}$ of the spectrum analyzer used, drops to below 50 kHz in the locking range, as shown in the inset in Fig. 2 (a). The resolution of 50kHz was chosen for faster data processing and analysis. However, the real linewidth inside the locking range decreases down to the linewidth of the external source $\approx$ 1Hz. In Fig. 2 (b) we show the corresponding Arnold tongue $P_{source}$ vs $I_{DC}$. The locking range expands with increasing RF power $P_{source}$. The corresponding instantaneous phase dynamics is shown in Fig. 2 (c) and the phase noise is shown in Fig. 2 (d) for an RF power of $-3dBm$ and three different values of detuning. (1) For large detuning beyond the locking range, the oscillator is unlocked and the phase is free, leading to a random walk of the instantaneous phase, which is reflected in large phase fluctuations in the time domain and a $1/f^2$ dependence in the phase noise (red curves,  Figs. 2 (c,d)). (2) At detuning values close to the locking range (red area in the Arnold tongue), the phase noise starts to flatten off below a roll-off frequency $f_{roll-off}$ (see Fig. 2 (d)). The instantaneous phase is partially locked but suffers from phase slips which is reflected in the increasing phase noise with at a slope of $1/f^2$ at offset frequencies below $f_{corner}$ (blue curves). (3) For zero detuning and detuning within approximately $60\%$ of the full locking range, there are no phase slips visible within the time window of basic characterization measurements (1ms). The phase noise becomes flat in the range of offset frequencies below $f_{roll-off}$ (black curves). It is in this area of the Arnold tongue with zero phase slips, for which phase modulation has to be carried out. To demonstrate binary phase shift keying (BPSK) we inject a time-varying modulation signal $V_{mod}$ in the form of a square wave pattern to the STNO for a given power $P_{source}=-3dBm$ of the injection locking external source and a given DC current. The amplitude of the modulation signal was chosen such that the detuning remains within the range free of phase slips.
\clearpage
\begin{figure}
    \centering
    \includegraphics[width=0.7\linewidth]{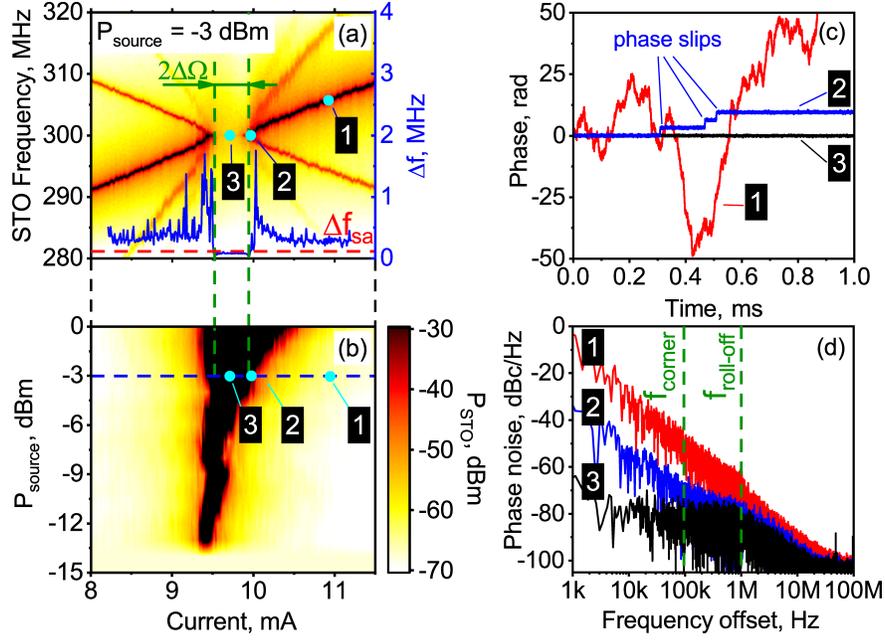}\\
    \caption{\textbf{\boldmath STNO's signal characteristics at $P_{source} = -3dB$ and $f_{source}=600MHz$ for different values of current:} 1 -  unlocked free-running STNO ($I_{STNO}= 11 mA$); 2 - partially locked STNO ($I_{STNO}= 10 mA$) with phase slips; 3 - injection locked STNO  ($I_{STNO}= 9.75 mA$). \textbf{(a)} State diagram. Corresponding linewidth (blue curve in inset). Red dashed line limits the resolution of spectrum analyzer. \textbf{(b)} Arnold's tongue demonstrating locking area in parameters of $P_{source}$(peak power) and $I_{STNO}$. The injection locking range at $P_{source} = -3dB$ highlighted by dashed blue line and delimited by vertical dashed green lines. \textbf{(c)} Instantaneous phase time traces. \textbf{(d)} Phase noise plot}
    \label{fig2}
\end{figure}

Similar to the phase noise analysis of the free running and injection-locked STNO (Fig.~2) we extract the instantaneous phase of the modulated STNO signal using a Hilbert transform technique. The data was obtained for $2f(N=2)$ and $f/2(N=1/2)$ locking regimes. The locking regime at $1f$ frequency has not been used since the STNO is a resistive one-port device and, consequently, it is impossible to separate the signal coming from the STNO and the external source. The time traces of the instantaneous phase are shown in Fig.~3 for different amplitudes $V_{mod}$ and for different modulation frequencies $f_{mod}$. Light blue lines represent the modulation signal $V_{mod}$ for $2f$ locking and red lines correspond to $V_{mod}$ for $f/2$ locking. Instantaneous phases are shown in dark blue ($2f$ synchronization) and pink ($f/2$ synchronization). In Fig.~3(a) for $2f$ synchronization at $f_{mod}=500kHz$, when the peak-to-peak amplitude of the square modulation signal $V_{mod}$ is $6 mV$, the shift of the instantaneous STNO phase $\Delta\Psi|_{STNO}$ is around $\approx0.5rad$. For $f/2$ synchronization at $V_{mod} = 4 mV$, the phase shift is $\Delta\Psi|_{STNO}\approx1.1rad$.  

\begin{figure}
    \centering
    \includegraphics[width=0.95\linewidth]{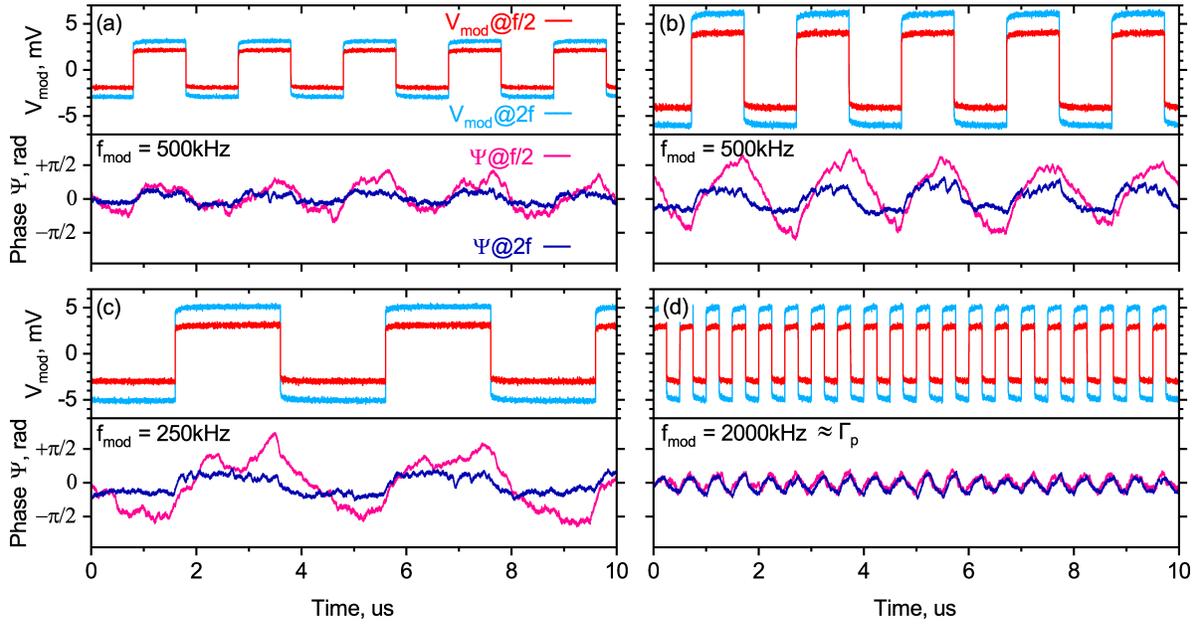}\\
    \caption{\textbf{\boldmath Time traces of the instantaneous phase and modulation signal in BPSK regime for $2f$ and $f/2$ locking frequencies.  $V_{mod}$ modulation amplitude: (a)} - $4mV@f/2$ and $6mV@2f$, $f_{mod}=500kHz$, \textbf{(b)} - $8mV@f/2$ and $12mV@2f$, $f_{mod}=500kHz$; \textbf{(c)} - $f_{mod}=250kHz$, $V_{mod}=6mV@f/2$ and $10mV@2f$, \textbf{(d)} - $f_{mod}=2MHz$, $V_{mod}=6mV@f/2$ and $10mV@2f$}
    \label{figTraces}
\end{figure}

The shift $\Delta\Psi|_{STNO}$ reaches $\approx0.9rad$ which is $\approx60\%$ of $\Delta\Psi_{max}$ for $2f$ locking and $\approx2.1rad$ which is $\approx70\%$ of $\Delta\Psi_{max}|_{STNO}$ for $f/2$ locking on increasing the modulation signal to $V_{mod}=12mV$ and $8mV$ respectively (see Fig.~\ref{figTraces}~(b)). This corresponds to a total change of current flowing through the STNO $2{\delta I}=V_{mod}/R_{STNO}\approx0.3mA$ for $2f$ locking and $2{\delta I}=V_{mod}/R_{STNO}\approx0.2mA$ for $f/2$ locking which corresponds to $60\%$ of the locking ranges ${2\delta\Omega(I)}$ at $2f$ and $f/2$. The signal-to-noise ratio (SNR) improves with increasing $V_{mod}$, Fig.~\ref{figTraces}~(b). However when increasing the amplitude of the square wave pattern above  $\approx60-70\%$ of the locking range $2{\delta I}$ the BPSK between well-defined levels is no more possible. Random fluctuations cause the instantaneous frequency to go beyond $60\%$ of the locking range ${2\Delta\Omega(I)}$, which leads to slips of the instantaneous phase within the measurement time window. Since phase slips are stochastic it will lead to the rise of bit error rate (BER). Thus, the maximal effective range of phase shifts is limited to $\approx60\%$ of the theoretical $\Delta\Psi_{max}|_{STNO}$. Another limitation appears at high frequency of modulation: when at constant amplitude of the modulation signal $V_{mod}$ the modulation frequency is increased to $f_{mod}=2000kHz$ the amplitude of the instantaneous phase shift drops to $\Delta\Psi\approx0.6rad$ (see Fig.~\ref{figTraces}~(d)) for both $2f$ and $f/2$ locking even though at $f_{mod}=250kHz$ (see Fig.~\ref{figTraces}~(c)) it is close to $\approx60\%$ of the theoretical limits. This is caused by the inertia-like behavior in the STNO in response to an external source and is reflected in the amplitude-relaxation frequency ${\Gamma}_p$ above which the influence of the external signal on the STNO becomes weaker and the STNO does not respond to external signals. ${\Gamma}_p$ is usually different for each synchronization regime \cite{Lebrun2015}. In our experiment ${\Gamma}_p$ is lower for $f/2$ locking than for $2f$ locking. It leads to a higher reduction of $\Delta\Psi|_{STNO}$ for $f/2$ than at $2f$ locking at the modulation frequency $f_{mod}=2MHz$.

\begin{figure} \label{figVoice}
    \centering
    \includegraphics[width=0.7\linewidth]{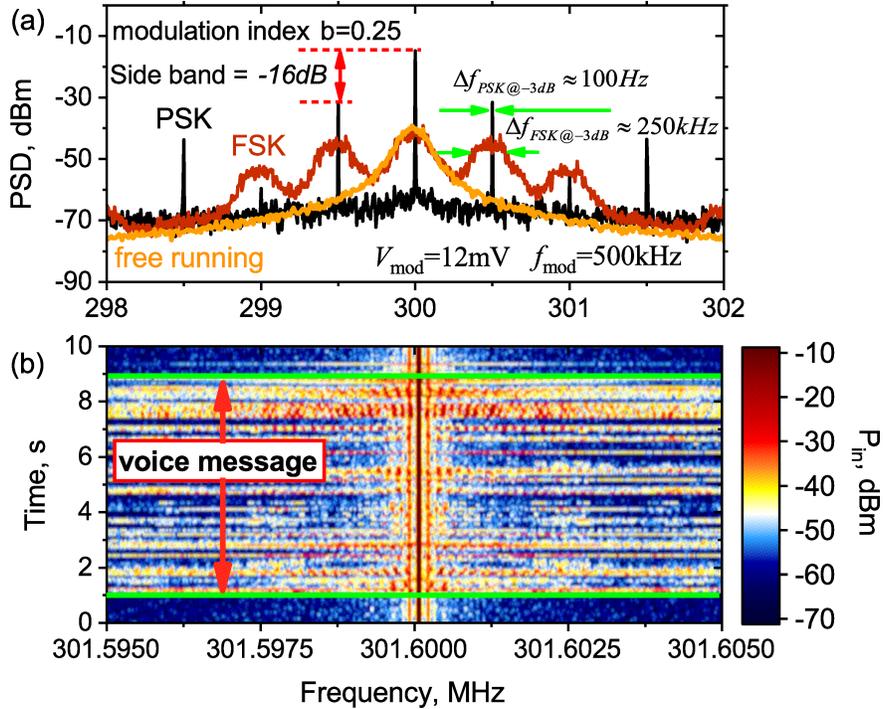}\\
    \caption{\textbf{(a)} - PSDs of  a free-running STNO (green curve), FSK signal (blue curve), BPSK signal in injection locking regime at  $P_{source} = -3dB$. Both modulated signals are obtained at the amplitude of modulation signal  $V_{mod} = 6mV$. Signal peak amplitude to noise floor ratio improves by a factor of $30dB$. \textbf{(b)} - Spectrogram of voice transmission. A voice message transcription: "Demonstration of voice transmission using phase modulation of spin torque nano-oscillators"}
\end{figure}

A complementary technique to characterize modulation properties of the STNO signal is spectrum analysis. The modulation rate, modulation index and the carrier linewidth can be easily deduced from the spectrum. In Fig. 4 (a) we compare the spectra for BPSK (black line), FSK (brown line, only $V_{mod}$ used, and zero  $P_{source}$) and the free running oscillations (orange line, zero $V_{mod}$ and zero $P_{source}$). The FSK and BPSK signals are characterized by multiple sidebands. In the FSK regime the instability of the carrier frequency translates to the modulation sidebands causing their broadening and, therefore, will degrade the quality of the demodulated signal. In the PSK regime the linewidth of modulation sidebands is defined by the stability of the external source and the modulation signal itself. With this it is possible to use modulation indices $\beta \leqslant 1$, while for the FSK scheme using STNOs \cite{Manfrini2011} it was necessary to use large modulation indices $\beta >> 1$ to be able to distinguish after demodulation between the actual signal and the fluctuations translated from the carrier. The disadvantage of large modulation indices $\beta >> 1$ is that the FSK occupied bandwidth $F_{occup}$ equals twice the peak frequency deviation $\Delta f$ of the frequency excursions $F_{occup}=2(\beta+1)f_{mod}\approx 2 \beta f_{mod}=2 \Delta f$ and is independent of the modulation frequency. It significantly reduces the number of frequency separated channels and hence the performance. The proposed PSK technique provides the carrier with narrow linewidth which allows to use narrow-band modulation protocols with modulation indices starting from $\approx0.1$. The spectrum for the BPSK signal under $V_{mod}=4mV$ for $2f$ locking is shown in Fig. 4. The amplitude of the first modulation harmonic is $16dB$ lower than the carrier. There are theoretical calculations of the dependence between the modulation index and relative sideband amplitudes for square wave modulation \cite{Stocklin1973}. For the first relative sideband amplitude of $-16dB$ the modulation index equals to $\beta=0.24rad$ which corresponds to one half of the full amplitude of phase shifts $\Delta\Psi|_{STNO}/2=0.22rad$. The good agreement between results from two different techniques confirms the correctness and the precision of the measurements. 

\begin{figure}
    \centering
    \includegraphics[width=0.8\linewidth]{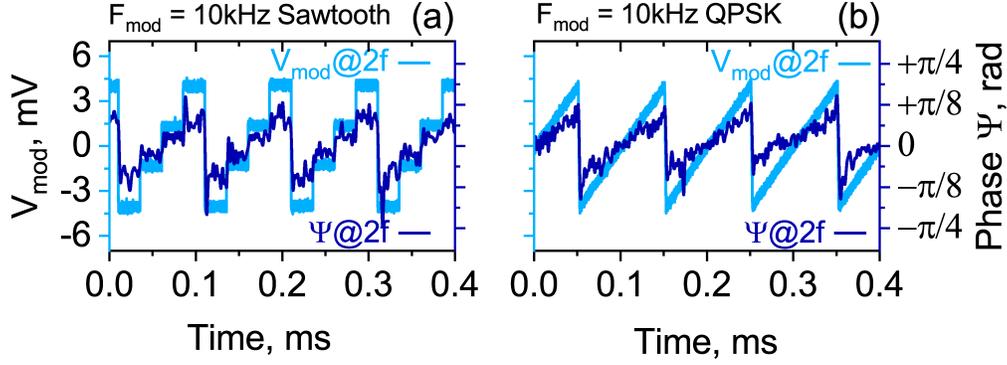}\\
    \caption{\textbf{\boldmath Time traces of the instantaneous phase and modulation signal in QPSK (a) and analog PM (b) regime.} $2f$ locking regime. $f_{mod}=10kHz$.}
    \label{figQPSK}
\end{figure}

The results of Figs.~3 and 4 clearly demonstrate the BPSK scheme using STNOs where the phase follows the input signal, in the limits discussed above. Since the dependence of the phase detuning is a continuous function of frequency detuning (eq.~1), more phase states can be used to decode data, thus increasing the data transmission capacity and channel efficiency. Here we demonstrate 4-level QPSK and continuous analog PM. While for demonstration of BPSK the SNR can be low to distinguish between two phase states, for demonstration of QPSK and analog PM the SNR has to be increased. Therefore, a low modulation rate of 10kHz was chosen and a low pass numerical filter with cut-off frequency of 300kHz was applied to the instantaneous phase signal to filter out phase noise above 30th harmonic. The high number of harmonics allows phase modulation with a complex shape signal. The amplitude of the modulation signal $V_{mod}$ was chosen to be 8mV to avoid phase slips on a millisecond timescale. It limits the maximum amplitude of the STNO phase shifts to $\Delta\Psi\approx0.7rad$. In Fig.~5~(a) and (b) the QPSK modulation with four distinguishable levels and analog PM are shown correspondingly. The phase change in analog PM is continuous and close to linear. The technique, therefore, allows to transmit analog information such as voice, analog sensors readings and etc. The frequency stability and the power of the PSK and PM signals generated by an injection-locked STNO are high enough to use conventional receivers such as FM- and TV-tuners. To demonstrate this, we have setup a benchtop transmitter - receiver system, where the transmitter uses the setup of Fig.1 and where the amplifier output is connected to an antenna and the modulation source is replaced by a microphone with an audio amplifier. At the receiver side the signal is captured by an antenna connected to a conventional software defined radio (SDR) receiver. The distance between the two antennae is 10m. More details can be found in the Supplementary. The SDR receiver is controlled by a PC program that displays the received signal as a frequency-time spectrogram to visualize the transmitted signals and produces an audible decoded signal. An example is shown in Fig. 4b. A real-time video can be found online\cite{supplvideo} with the voice transmission in the form of frequency-time spectrogram with an audible soundtrack. When there is no input to the microphone, only the carrier is seen. After 1s the microphone is switched on and the voice transmission starts, generating sidebands on the carrier that are clearly seen.

In summary we have demonstrated the concept of digital phase shift keying and analog phase modulation in MTJ based STNO with vortex configuration. The maximum data rate of 4 Mbit/s was achieved for both $2f$ and $f/2$ synchronization regimes and found to be limited by the amplitude relaxation rate. The phase shift is directly proportional to the modulation signal $V_{mod}$. The concept was extended to demonstrate voice transmission using analog modulation as well as quadrature phase shift keying (QPSK) by injecting a modulation signal with four distinct levels. The demonstration of PSK presented here has been done for injection locking at $2f$ and $f/2$ since for injection locking at $f$ the signal of the STNO and the external source cannot be distinguished. While injection locking at $2f$ limits the theoretical phase shift $\Delta\Psi_{max}$ to $\pi/2$, the advantage of sub-harmonic locking is that theoretically the full range of phase shift $\Delta\Psi_{max}=\pi$ can be used.

Here we used the injection of current generated by an arbitrary waveform voltage generator as the modulating signal. As shown previously \cite{Purbawati2016EnhancedFM}, this limits the maximum possible modulation rate to the amplitude relaxation frequency - ${\Gamma}_p$. For vortex devices ${\Gamma}_p$ lies on the order of several MHz. Using a microstrip line to generate a modulation RF magnetic field one can improve the maximum possible modulation rate above 2MHz for vortex devices. However, vortex devices have been selected here merely for the demonstration, since they show perfect locking. For uniform in-plane devices, which have ${\Gamma}_p$ in the range of 100-200MHz \cite{Purbawati2016EnhancedFM}, FSK was demonstrated using current modulation with data transmission rates up to 400Mbit/s \cite{Ruiz-Calaforra2017}. Hence, in principle much higher modulation rates can be achieved with the proposed PSK technique for STNOs operating at higher frequencies.

The proposed technique presented in this paper can be applied and is best suited for any type of nano-scale auto-oscillators where the phase noise is high due to nanoscale dimension and where it can be sufficiently improved with an external synchronization signal. As an outlook, we note that the use of many, either external or internal, RF sources for every channel in communication systems leads to consumption of area and complicates the transmitter design. As a development of the technique, we are going to exclude the external RF sources and use separately controlled PLL-stabilized STNOs \cite{Kreissig2017} working at different frequency channels. Such STNO-PLLs can thus be considered as a stable RF reference source with a stable reference phase. Injection locking of a second STNO to such a phase stabilized STNO-PLL will provide a route to implement phase modulation as a compact and effective communication scheme for integrated circuits(ICs).

\begin{methods}
\subsection{Sample fabrication.} The composition of the vortex MTJs used in our experiments is the following:  5 Ta/ 50 CuN/ 5 Ta/ 50 CuN/ 5 Ta/ 5 Ru/ 6 IrMn/ 2.6 CoFe30/ 0.85 Ru/ 1.8 CoFe40B20/ MgO/ 2.0 CoFe40B20/ 0.2 Ta/ 7 NiFe/ 10 Ta/ 30 CuN/ 7 Ru [thicknesses in nm]. The materials were sputter deposited and nanofabricated at INL using a Singulus Rotaris machine and ion beam and optical etching techniques. Electrical measurements were performed using a 50-Ohm matched nano-probe station. The sample was placed on top of a permanent magnet using a positioning stage to provide a field of ${1-3kOe}$ at a small out-of-plane angle of ${1-5}$ degrees. While different devices have been measured, with similar performances, results for two devices of 370 nm diameter with similar dynamical and static characteristics (${R_{STNO}\approx40 Ohm}$, ${TMR\approx150\%}$ and ${RA\approx4.5{\Omega\mu}m^2}$) are presented in the paper. 

\subsection{Data post-processing.}
To better asses the dynamics of the instantaneous phase $\phi(t)$, we registered the time traces $x(t)$ of the voltage output signal from which the phase fluctuations are extracted using the Hilbert transform \cite{Quinsat2014, Lebrun2015, GrimaldiPhysRevB.89.104404, Ruiz-Calaforra2017, Purbawati2016EnhancedFM}:
\begin{equation}
    \phi(t)=arctan(\frac{HT[x(t)]}{x(t)})
\end{equation}
where HT is the Hilbert transform. The data shown in Fig.~2~(c,d) corresponds to the instantaneous phase in relation to the locking frequency for injection locked and partially locked states: 
\begin{equation}
    \delta\phi(t)=\phi(t)-2\pi f_{source}t/N
\end{equation}
and to the phase in relation to the average STNO frequency for unlocked free-running state:
\begin{equation}
    \delta\phi(t)=\phi(t)-2\pi f_{STNO}t
\end{equation}
The analysis of the phase noise figure was done by taking the power spectral density of the instantaneous phase fluctuations $\delta\phi(t)$  (\ref{eq:1},\ref{eq:1}):
\begin{equation}
    PSD_{\delta\phi}= FFT^2(\delta\phi(t))
\end{equation}

The Arnold tongue was assembled from data of state diagrams at different $P_{source}$. 
From each state diagram a line at $f_{source}$ was taken to assemble the Arnold tongue contour plot, where the y-axis is $P_{source}$, color-axis is $P_{STNO}$.   The accuracy of this method is limited by the free running linewidth. It also doesn't distinguish between general and complete synchronization regimes. However, it is accurate enough to determine the center of the locking range and the approximate maximum amplitude of the modulation signal $V_{mod}=I_{mod}R$ which has to be applied to achieve the maximum phase detuning $\Delta\Psi_{max}$. 

For experiments with PM and PSK the value of $P_{source}$ was chosen in the linear part of the Arnold tongue to avoid nonlinear effects in frequency and phase detuning. 
\end{methods}



\bibliographystyle{naturemag}
\bibliography{references.bib}

\begin{thebibliography}{10}
\expandafter\ifx\csname url\endcsname\relax
  \def\url#1{\texttt{#1}}\fi
\expandafter\ifx\csname urlprefix\endcsname\relax\def\urlprefix{URL }\fi
\providecommand{\bibinfo}[2]{#2}
\providecommand{\eprint}[2][]{\url{#2}}

\bibitem{YangWSN2013}
\bibinfo{author}{Yang, S.-H.}
\newblock \emph{\bibinfo{title}{Wireless Sensor Networks: Principles, Design
  and Applications}} (\bibinfo{publisher}{Springer Publishing Company,
  Incorporated}, \bibinfo{year}{2013}).

\bibitem{Uckelmann2011}
\bibinfo{author}{Uckelmann, D.}, \bibinfo{author}{Harrison, M.} \&
  \bibinfo{author}{Michahelles, F.}
\newblock \emph{\bibinfo{title}{Architecting the internet of things}}
  (\bibinfo{publisher}{Springer Science \& Business Media},
  \bibinfo{year}{2011}).

\bibitem{Dumas2014RecentAdv}
\bibinfo{author}{Dumas, R.~K.} \emph{et~al.}
\newblock \bibinfo{title}{Recent advances in nanocontact spin-torque
  oscillators}.
\newblock \emph{\bibinfo{journal}{IEEE transactions on magnetics}}
  \textbf{\bibinfo{volume}{50}}, \bibinfo{pages}{1--7} (\bibinfo{year}{2014}).

\bibitem{Chen2016SpinTorqueAS}
\bibinfo{author}{Chen, T.} \emph{et~al.}
\newblock \bibinfo{title}{Spin-torque and spin-hall nano-oscillators}.
\newblock \emph{\bibinfo{journal}{Proceedings of the IEEE}}
  \textbf{\bibinfo{volume}{104}}, \bibinfo{pages}{1919--1945}
  (\bibinfo{year}{2016}).

\bibitem{Choi2014}
\bibinfo{author}{{H. S. Choi, S. Y. Kang, S. J. Cho, I.-Y. Oh, M. Shin, H.
  Park, C. Jang, B.-C. Min, S.-I. Kim, S.-Y. Park et al.}}
\newblock \emph{\bibinfo{journal}{Sci. Rep.}} \textbf{\bibinfo{volume}{4}},
  \bibinfo{pages}{5486} (\bibinfo{year}{2014}).

\bibitem{Oh2014}
\bibinfo{author}{{Oh}, I.}, \bibinfo{author}{{Park}, S.},
  \bibinfo{author}{{Kang}, D.} \& \bibinfo{author}{{Park}, C.~S.}
\newblock \bibinfo{title}{Wireless spintronics modulation with a spin torque
  nano-oscillator (stno) array}.
\newblock \emph{\bibinfo{journal}{IEEE Microwave and Wireless Components
  Letters}} \textbf{\bibinfo{volume}{24}}, \bibinfo{pages}{502--504}
  (\bibinfo{year}{2014}).

\bibitem{Sharma2015ComSys}
\bibinfo{author}{Sharma, R.} \emph{et~al.}
\newblock \bibinfo{title}{Modulation rate study in a spin-torque
  oscillator-based wireless communication system}.
\newblock \emph{\bibinfo{journal}{IEEE Transactions on Magnetics}}
  \textbf{\bibinfo{volume}{51}}, \bibinfo{pages}{1--4} (\bibinfo{year}{2015}).

\bibitem{Purbawati2016EnhancedFM}
\bibinfo{author}{Purbawati, A.}, \bibinfo{author}{Garcia-Sanchez, F.},
  \bibinfo{author}{Buda-Prejbeanu, L.} \& \bibinfo{author}{Ebels, U.}
\newblock \bibinfo{title}{Enhanced modulation rates via field modulation in
  spin torque nano-oscillators}.
\newblock \emph{\bibinfo{journal}{Applied Physics Letters}}
  \textbf{\bibinfo{volume}{108}}, \bibinfo{pages}{122402}
  (\bibinfo{year}{2016}).

\bibitem{Slonczewski1996}
\bibinfo{author}{{J. C. Slonczewski}}.
\newblock \emph{\bibinfo{journal}{Magn. Magn. Mater.}}
  \textbf{\bibinfo{volume}{159}}, \bibinfo{pages}{L1–L7}
  (\bibinfo{year}{1996}).

\bibitem{Lebrun2015}
\bibinfo{author}{{R. Lebrun et al.}}
\newblock \emph{\bibinfo{journal}{Phys. Rev. Lett.}}
  \textbf{\bibinfo{volume}{115}}, \bibinfo{pages}{017201}
  (\bibinfo{year}{2015}).

\bibitem{Pufall2005FM}
\bibinfo{author}{Pufall, M.~R.}, \bibinfo{author}{Rippard, W.~H.},
  \bibinfo{author}{Kaka, S.}, \bibinfo{author}{Silva, T.~J.} \&
  \bibinfo{author}{Russek, S.~E.}
\newblock \bibinfo{title}{Frequency modulation of spin-transfer oscillators}.
\newblock \emph{\bibinfo{journal}{Applied Physics Letters}}
  \textbf{\bibinfo{volume}{86}}, \bibinfo{pages}{082506}
  (\bibinfo{year}{2005}).
\newblock \urlprefix\url{https://doi.org/10.1063/1.1875762}.
\newblock \eprint{https://doi.org/10.1063/1.1875762}.

\bibitem{Muduli2010NonlinFMandAM}
\bibinfo{author}{Muduli, P.~K.} \emph{et~al.}
\newblock \bibinfo{title}{Nonlinear frequency and amplitude modulation of a
  nanocontact-based spin-torque oscillator}.
\newblock \emph{\bibinfo{journal}{Phys. Rev. B}} \textbf{\bibinfo{volume}{81}},
  \bibinfo{pages}{140408} (\bibinfo{year}{2010}).
\newblock \urlprefix\url{https://link.aps.org/doi/10.1103/PhysRevB.81.140408}.

\bibitem{Manfrini2011}
\bibinfo{author}{Manfrini, M.} \emph{et~al.}
\newblock \bibinfo{title}{Frequency shift keying in vortex-based spin torque
  oscillators}.
\newblock \emph{\bibinfo{journal}{Journal of Applied Physics}}
  \textbf{\bibinfo{volume}{109}}, \bibinfo{pages}{083940}
  (\bibinfo{year}{2011}).

\bibitem{Sharma2017SBModulation}
\bibinfo{author}{Sharma, R.} \emph{et~al.}
\newblock \bibinfo{title}{A high-speed single sideband generator using a
  magnetic tunnel junction spin torque nano-oscillator}.
\newblock \emph{\bibinfo{journal}{Scientific reports}}
  \textbf{\bibinfo{volume}{7}}, \bibinfo{pages}{13422} (\bibinfo{year}{2017}).

\bibitem{Ruiz-Calaforra2017}
\bibinfo{author}{{A. Ruiz-Calaforra, et al.}}
\newblock \emph{\bibinfo{journal}{Applied Physics Letters}}
  \textbf{\bibinfo{volume}{111}}, \bibinfo{pages}{082401}
  (\bibinfo{year}{2017}).

\bibitem{Zahedinejad2017CurrentAMFM}
\bibinfo{author}{Zahedinejad, M.} \emph{et~al.}
\newblock \bibinfo{title}{Current modulation of nanoconstriction spin-hall
  nano-oscillators}.
\newblock \emph{\bibinfo{journal}{IEEE Magnetics Letters}}
  \textbf{\bibinfo{volume}{8}}, \bibinfo{pages}{1--4} (\bibinfo{year}{2017}).

\bibitem{Haykin1988}
\bibinfo{author}{Haykin, S.}
\newblock \emph{\bibinfo{title}{Digital Communications}}
  (\bibinfo{publisher}{John Wiley \& Sons, Toronto, Canada},
  \bibinfo{year}{1988}).

\bibitem{Anderson2013PSK}
\bibinfo{author}{Anderson, J.~B.}, \bibinfo{author}{Aulin, T.} \&
  \bibinfo{author}{Sundberg, C.-E.}
\newblock \emph{\bibinfo{title}{Digital phase modulation}}
  (\bibinfo{publisher}{Springer Science \& Business Media},
  \bibinfo{year}{2013}).

\bibitem{Tewari2016}
\bibinfo{author}{{J. Tewari, H. M. Singh}}.
\newblock \emph{\bibinfo{journal}{J. Inn. Res. in Comp. and Comm. Eng.}}
  \textbf{\bibinfo{volume}{4}}, \bibinfo{pages}{13425--13431}
  (\bibinfo{year}{2016}).

\bibitem{IEEE2003}
\bibinfo{title}{{Wireless LAN Medium Access Control (MAC) and Physical Layer
  (PHY) specifications}}.
\newblock \emph{\bibinfo{journal}{IEEE Std. 802.11g-2003}}
  (\bibinfo{year}{2003}).

\bibitem{Kreissig2017}
\bibinfo{author}{{M. Kreißig et al.}}
\newblock \emph{\bibinfo{journal}{IEEE 60th International Midwest Symposium on
  Circuits and Systems (MWSCAS)}} \textbf{\bibinfo{volume}{0}},
  \bibinfo{pages}{910--913} (\bibinfo{year}{2009}).

\bibitem{Tamaru2016}
\bibinfo{author}{{S. Tamaru, et al.}}
\newblock \emph{\bibinfo{journal}{Applied Physics Express}}
  \textbf{\bibinfo{volume}{5}}, \bibinfo{pages}{053005} (\bibinfo{year}{2016}).

\bibitem{Adler1946}
\bibinfo{author}{Adler, R.}
\newblock \bibinfo{title}{A study of locking phenomena in oscillators}.
\newblock \emph{\bibinfo{journal}{Proceedings of the IRE}}
  \textbf{\bibinfo{volume}{34}}, \bibinfo{pages}{351--357}
  (\bibinfo{year}{1946}).

\bibitem{Kiselev2003}
\bibinfo{author}{{S. I. Kiselev, J. C. Sankey, I. N. Krivorotov, N. C. Emley,
  R. J. Schoelkopf, R. A. Buhrman, and D. C. Ralph}}.
\newblock \emph{\bibinfo{journal}{Nature}} \textbf{\bibinfo{volume}{425}},
  \bibinfo{pages}{380} (\bibinfo{year}{2003}).

\bibitem{Grollier2006}
\bibinfo{author}{{J. Grollier, V. Cros, and A. Fert}}.
\newblock \emph{\bibinfo{journal}{Phys. Rev. B}} \textbf{\bibinfo{volume}{73}},
  \bibinfo{pages}{060409} (\bibinfo{year}{2006}).

\bibitem{Tiberkevich2009}
\bibinfo{author}{{V. Tiberkevich, A. Slavin, E. Bankowski, and G. Gerhart}}.
\newblock \emph{\bibinfo{journal}{Appl. Phys. Lett.}}
  \textbf{\bibinfo{volume}{95}}, \bibinfo{pages}{262505}
  (\bibinfo{year}{2009}).

\bibitem{Demidov2014ExtSynchro}
\bibinfo{author}{Demidov, V.} \emph{et~al.}
\newblock \bibinfo{title}{Synchronization of spin hall nano-oscillators to
  external microwave signals}.
\newblock \emph{\bibinfo{journal}{Nature communications}}
  \textbf{\bibinfo{volume}{5}}, \bibinfo{pages}{3179} (\bibinfo{year}{2014}).

\bibitem{Tortarolo2018}
\bibinfo{author}{{M. Tortarolo et al.}}
\newblock \emph{\bibinfo{journal}{Scientific Reports}}
  \textbf{\bibinfo{volume}{8}}, \bibinfo{pages}{1728} (\bibinfo{year}{2018}).

\bibitem{Zhou2007}
\bibinfo{author}{{Y. Zhou, et al.}}
\newblock \emph{\bibinfo{journal}{Journal of Applied Physics.}}
  \textbf{\bibinfo{volume}{101}}, \bibinfo{pages}{09A510}
  (\bibinfo{year}{2007}).

\bibitem{Rippard2005}
\bibinfo{author}{{W. Rippard, et al.}}
\newblock \emph{\bibinfo{journal}{Phys. Rev. Lett.}}
  \textbf{\bibinfo{volume}{95}}, \bibinfo{pages}{067203}
  (\bibinfo{year}{2005}).

\bibitem{Rosenblum2001}
\bibinfo{author}{M.~Rosenblum, J. K. C.~S., A.~Pikovsky} \&
  \bibinfo{author}{Tass, P.~A.}
\newblock \bibinfo{title}{Phase synchronization: From theory to data analysis}.
\newblock \bibinfo{pages}{279–--321} (\bibinfo{publisher}{Elsevier,
  Amsterdam}, \bibinfo{year}{2001}).

\bibitem{Slavin2009}
\bibinfo{author}{{A. N. Slavin and V. Tiberkevich}}.
\newblock \emph{\bibinfo{journal}{Trans. Magn.}} \textbf{\bibinfo{volume}{45}},
  \bibinfo{pages}{1875} (\bibinfo{year}{2009}).

\bibitem{Balanov2009}
\bibinfo{author}{Balanov, A.}, \bibinfo{author}{Janson, N.},
  \bibinfo{author}{Postnov, D.} \& \bibinfo{author}{Sosnovtseva, O.}
\newblock \emph{\bibinfo{title}{Synchronization: From Simple to Complex}}
  (\bibinfo{publisher}{Springer}, \bibinfo{year}{2009}).

\bibitem{Pikovsky2001}
\bibinfo{author}{Pikovsky, A.}, \bibinfo{author}{Rosenblum, M.} \&
  \bibinfo{author}{Kurths, J.}
\newblock \emph{\bibinfo{title}{Synchronization : a universal concept in
  nonlinear sciences}} (\bibinfo{publisher}{Cambridge University Press,
  Cambridge}, \bibinfo{year}{2001}).

\bibitem{Osipov2007}
\bibinfo{author}{Osipov, G.~V.}, \bibinfo{author}{Kurths, J.} \&
  \bibinfo{author}{Zhou, C.}
\newblock \emph{\bibinfo{title}{Synchronization in oscillatory networks}}
  (\bibinfo{publisher}{Springer Science \& Business Media},
  \bibinfo{year}{2007}).

\bibitem{Razavi2004}
\bibinfo{author}{Razavi, B.}
\newblock \bibinfo{title}{A study of injection locking and pulling in
  oscillators}.
\newblock \emph{\bibinfo{journal}{IEEE Journal of Solid-State Circuits}}
  \textbf{\bibinfo{volume}{39}}, \bibinfo{pages}{1415--1424}
  (\bibinfo{year}{2004}).

\bibitem{Urazhdin2010}
\bibinfo{author}{{S. Urazhdin, P. Tabor, V. Tiberkevich and A. Slavin}}.
\newblock \emph{\bibinfo{journal}{Phys. Rev. Lett.}}
  \textbf{\bibinfo{volume}{105}}, \bibinfo{pages}{104101}
  (\bibinfo{year}{2010}).

\bibitem{Quinsat2014}
\bibinfo{author}{{M. Quinsat, F. Garcia-Sanchez, A. S. Jenkins, A. Zeltser, J.
  A. Katine, A. N. Slavin, L. D. Buda-Prejbeanu, B. Dieny, M.-C. Cyrille, and
  U. Ebels}}.
\newblock \emph{\bibinfo{journal}{Applied Physics Letters}}
  \textbf{\bibinfo{volume}{105}}, \bibinfo{pages}{152401}
  (\bibinfo{year}{2014}).

\bibitem{GrimaldiPhysRevB.89.104404}
\bibinfo{author}{Grimaldi, E.} \emph{et~al.}
\newblock \bibinfo{title}{Response to noise of a vortex based spin transfer
  nano-oscillator}.
\newblock \emph{\bibinfo{journal}{Phys. Rev. B}} \textbf{\bibinfo{volume}{89}},
  \bibinfo{pages}{104404} (\bibinfo{year}{2014}).
\newblock \urlprefix\url{https://link.aps.org/doi/10.1103/PhysRevB.89.104404}.

\bibitem{Kim2008}
\bibinfo{author}{{J.-V. Kim, V. Tiberkevich, et A. N. Slavin}}.
\newblock \emph{\bibinfo{journal}{Phys. Rev. Lett.}}
  \textbf{\bibinfo{volume}{100}}, \bibinfo{pages}{017207}
  (\bibinfo{year}{2016}).

\bibitem{Stocklin1973}
\bibinfo{author}{Stocklin, F.}
\newblock \bibinfo{title}{Relative sideband amplitudes versus modulation index
  for common functions using frequency and phase modulation}.
\newblock \bibinfo{type}{Tech. Rep.}, \bibinfo{institution}{Goddard Space
  Flight Center}, \bibinfo{address}{Greenbelt, Maryland, USA}
  (\bibinfo{year}{1973}).
\newblock \bibinfo{note}{An optional note}.

\bibitem{supplvideo}
\bibinfo{title}{{Video Demonstration of Voice Transmission using PM}}.
\newblock \bibinfo{howpublished}{\url{http://www.spintec.fr/}}.

\end{thebibliography}


\begin{thebibliography}{1}
\expandafter\ifx\csname url\endcsname\relax
  \def\url#1{\texttt{#1}}\fi
\expandafter\ifx\csname urlprefix\endcsname\relax\def\urlprefix{URL }\fi
\providecommand{\bibinfo}[2]{#2}
\providecommand{\eprint}[2][]{\url{#2}}

\bibitem{Rafael:R820T}
\bibinfo{organization}{{ Rafael Microelectronics, Inc.}}
\newblock \emph{\bibinfo{title}{R820T. High Performance Low Power Advanced
  Digital TV Silicon Tuner. Datasheet}} (\bibinfo{year}{2011}).

\end{thebibliography}


\begin{addendum}
 \item The authors acknowledge Dmitry Postnov for fruitful discussion on phase detuning of an externally synchronized oscillator. Financial support is acknowledged from the French space agency CNES, Enhanced Eurotalent program and from ERC MagiCal 669204.
 \item[Competing Interests] The authors declare that they have no
competing financial interests.
 \item[Correspondence] Correspondence and requests for materials
should be addressed to A.L.~(email: artem.litvinenko@cea.fr) and U.E.~(email: ursula.ebels@cea.fr)
 \item[Data availability] The data that support the plots within this paper and other findings of this study are available from the corresponding authors upon reasonable request. 
\end{addendum}

\end{document}